\documentclass[letterpaper,english, superscriptaddress,twocolumn,prl,letterpaper, nofootinbib]{revtex4}

\usepackage[]{fontenc}
\usepackage[latin9]{inputenc}
\usepackage{textcomp}
\usepackage{amsmath}
\usepackage{graphicx}
\usepackage{amssymb}

\usepackage{floatrow}    %% floatrow package-- it centers the content of floats by default.

\makeatletter

\usepackage{float}

\usepackage{babel}
\begin{document}
\title{Formation and observation of a quasi-two-dimensional $d_{xy}$ electron liquid in epitaxially stabilized Sr$_{2-x}$La$_{x}$TiO$_{4}$ thin films}

\author{Y.F. Nie}
\affiliation{National Laboratory of Solid State Microstructures and College of Engineering and Applied Sciences, Nanjing University, Nanjing, 210093, P.R. China}
\affiliation{Laboratory of Atomic and Solid State Physics, Department of Physics, Cornell University, Ithaca, New York 14853, USA}
\affiliation{Department of Materials Science and Engineering, Cornell University, Ithaca, New York 14853, USA}
\author{D. Di Sante}
\affiliation{Department of Physical and Chemical Sciences, University of L'Aquila, via Vetoio, I-67100 Coppito, L'Aquila, Italy}
\affiliation{CNR-SPIN, Via Vetoio, L'Aquila, I-67100, Italy}
\author{S. Chatterjee}
\affiliation{Laboratory of Atomic and Solid State Physics, Department of Physics, Cornell University, Ithaca, New York 14853, USA}
\author{P.D.C. King}
\affiliation{Laboratory of Atomic and Solid State Physics, Department of Physics, Cornell University, Ithaca, New York 14853, USA}
\affiliation{Kavli Institute at Cornell for Nanoscale Science, Ithaca, New York 14853, USA}
\author{M. Uchida}
\affiliation{Laboratory of Atomic and Solid State Physics, Department of Physics, Cornell University, Ithaca, New York 14853, USA}
\author{S. Ciuchi}
\affiliation{Department of Physical and Chemical Sciences, University of L'Aquila, via Vetoio, Coppito, L'Aquila, I-67100, Italy}
\affiliation{CNR-ISC, Via dei Taurini, Rome, I-00185, Italy}
\author{D.G. Schlom}  
\affiliation{Department of Materials Science and Engineering, Cornell University, Ithaca, New York 14853, USA}
\affiliation{Kavli Institute at Cornell for Nanoscale Science, Ithaca, New York 14853, USA}
\author{K.M. Shen}
\email[Author to whom correspondence should be addressed: ]{kmshen@cornell.edu}
\affiliation{Laboratory of Atomic and Solid State Physics, Department of Physics, Cornell University, Ithaca, New York 14853, USA}
\affiliation{Kavli Institute at Cornell for Nanoscale Science, Ithaca, New York 14853, USA}

%\date{\today}

%\doublespace % comment this out for single spacing

%%%%%%%%%%%%%%%%% abstract\section{Abstract}
\begin{abstract}
%A wide variety of emergent phenomena, including superconductivity and magnetism, have recently been discovered in two-dimensional electron liquids formed at titanate interfaces. The vast majority of these systems have been realized in perovskites such as SrTiO$_{3}$, but the Ruddlesden-Popper family (Sr$_{n+1}$Ti$_{n}$O$_{3n+1}$) offers another versatile platform for engineering interfaces. Here 

We report the formation and observation of an electron liquid in Sr$_{2-x}$La$_{x}$TiO$_4$, the quasi-two-dimensional counterpart of SrTiO$_3$, through reactive molecular-beam epitaxy and {\it in situ} angle-resolved photoemission spectroscopy. The lowest lying states are found to be comprised of Ti 3$d_{xy}$ orbitals, analogous to the LaAlO$_3$/SrTiO$_3$ interface and exhibit unusually broad features characterized by quantized energy levels and a reduced Luttinger volume. Using model calculations, we explain these characteristics through an interplay of disorder and electron-phonon coupling acting co-operatively at similar energy scales, which provides a possible mechanism for explaining the low free carrier concentrations observed at various oxide heterostructures such as the LaAlO$_3$/SrTiO$_3$ interface.
 
\end{abstract}
 
\maketitle

%%%%%%%%%%%%%%%%%  intro stuff 

%\textbf{Introduction}

Strontium titanate (SrTiO$_{3}$) is a  key building block in oxide electronics, particularly for the formation of two-dimensional electron liquids (2DELs) at SrTiO$_{3}$ interfaces and surfaces. For instance, the interface between LaAlO$_3$/SrTiO$_3$ ~\cite{ohtomo2004high}, exhibits exotic properties such as superconductivity and ferromagnetism. There have also been  reports of high temperature superconductivity at the monolayer FeSe / SrTiO$_{3}$ interface. At the LaAlO$_3$/SrTiO$_3$ interface or SrTiO$_{3}$ surface, the degeneracy of the $t_{2g}$ orbitals is lifted \cite{Cancellieri.PRB.89.121412, Berner.PRL.110.247601,king2014quasiparticle}, and the $d_{xy}$ states play an important role in the observed ferromagnetism ~\cite{lee2013titanium} and the superconducting phase diagram ~\cite{joshua2012universal, richter2013interface}. An alternate strategy to break the $t_{2g}$ degeneracy and realize a 2DEL is to interrupt the three-dimensional network of corner-sharing TiO$_{6}$ octahedra with a double SrO layer, resulting in Sr$_{2}$TiO$_{4}$, the quasi-two-dimensional $n$ = 1 end member of the Ruddlesden-Popper (RP) series Sr$_{n+1}$Ti$_{n}$O$_{3n+1}$, of which SrTiO$_{3}$ is the other three-dimensional, $n = \infty$ end member. Sr$_{2}$TiO$_{4}$ could also provide a natural platform for exploring exotic superconductivity, ferroelectricity, and magnetism in the titanates. Bulk single crystals of Sr$_{2}$TiO$_{4}$ have never been synthesized~\cite{Tilley1977293}, hindering the utility and understanding of this compound, and its electronic structure remains unexplored to date.

{\color{black}In this Letter, we report the synthesis and investigation of the electronic structure of lanthanum doped Sr$_{2-x}$La$_{x}$TiO$_{4}$ by a combination of oxide molecular-beam epitaxy (MBE) and  $in~situ$ angle-resolved photoemission spectroscopy (ARPES). We observe a single $d_{xy}$ band resulting in a circular Fermi surface near the Fermi energy, in addition to a number of unexpected features. The measured Luttinger volume is approximately a factor of 4 smaller than expected, and the spectra exhibit a series of energy levels with $\Delta =$ 93 meV separation, reminiscent of the phonon replicas recently reported in monolayer FeSe on SrTiO$_{3}$~\cite{lee2014}. We employ model calculations which cooperatively treat the effects of disorder and electron-phonon (el-ph) coupling to successfully explain these apparent discrepancies.

%We observe a quasi-localized electronic states showing unusual quantization features, which can be qualitatively interoperated by our theory calculation of a Anderson-Holstein model. 

%%%%%%%%%%%%%%%%%%%%%%%%%%%%%%%%%%%FIG 1A%%%%%%%%%%%%%%%%%%%%
\begin{figure*}
\begin{center}
\includegraphics{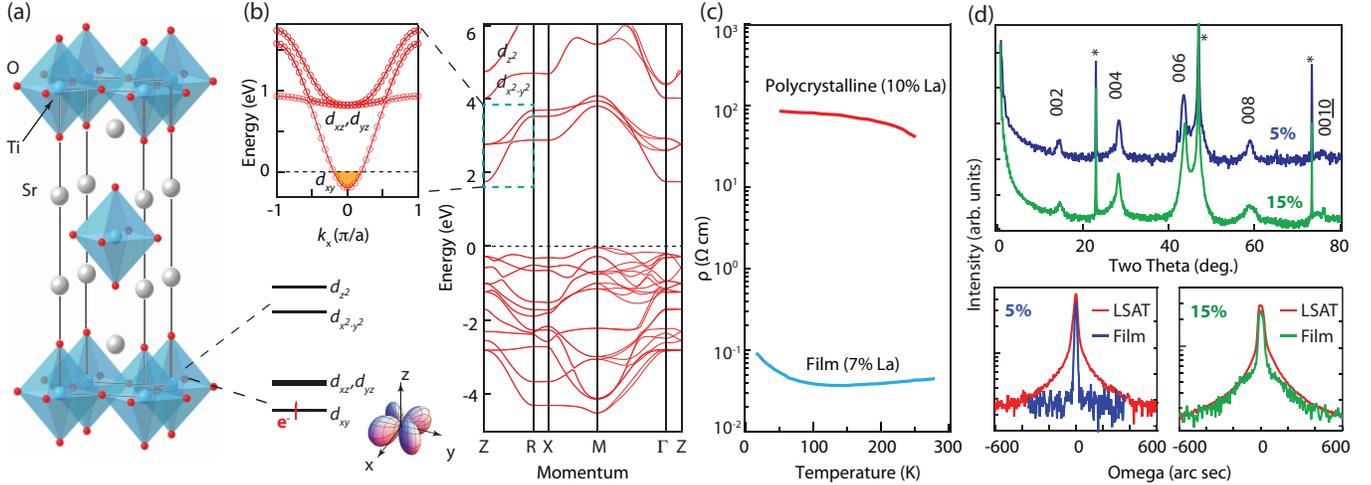}
\caption{(color online) (a) Diagram of the crystal structure of Sr$_2$TiO$_4$. (b) DFT band calculations of Sr$_{2}$TiO$_{4}$, showing the lowest unoccupied band of $d_\mathrm{xy}$ character. The left enlarged plot shows a rigid band shift of 5\% electron doping. (c) An epitaxial La-doped Sr$_2$TiO$_4$ thin film shows orders of magnitude lower resistivity than a bulk polycrystalline sample~\cite{Sugimoto1998315}.  (d) X-ray diffraction and rocking curves of the 006 diffraction peak of 5\% and 15\% La-doped Sr$_{2}$TiO$_{4}$ films. The X-ray diffraction scans are offset for clarity and the substrate peaks are marked with asterisks (*).
\label{fig:DFT}}
\end{center}
\end{figure*}

%%%%%%%%%%%%%%%%%%%%%%%%%%%%%%%%%%%FIG 1A%%%%%%%%%%%%%%%%%%%%

%%%%%%%%%%%%%%%% Results

%\textbf{Results}

%%%%%%%%%%%%%%%%% description of figure 1 (Sample characterizations)

Epitaxial La-doped Sr$_2$TiO$_4$ films with a thickness of 20 formula units were grown on (001) LSAT  [(LaAlO$_3$)$_{0.3}$ (SrAl$_{1/2}$Ta$_{1/2}$O$_3$)$_{0.7}$] substrates by MBE at 850$^{\circ}$ C in 10$^{-7}$ torr of O$_{2}$ using a shuttered growth technique~\cite{Haeni2000}. $in~situ$ ARPES measurements were taken at 20 K at an energy resolution of $\Delta E=12$ meV with a VG Scienta R4000 electron analyzer and He-I$\alpha$ photons ($h\nu$= 21.218 eV). Details of the sample growth,  DFT calculations, and the theoretical model are provided in the supplementary, which includes Refs. [12-15].%~\cite{Perdew1996,Giannozzi:2009hx,alvarez1985,mostofi2008wannier90}. 

Due to the layered crystal structure and an increase in the titanium
apical oxygen distance over the in-plane oxygens (2.1\%), the degeneracy
in Ti $t_{2g}$ orbitals is removed [Fig. 1(b)]. Similar to the high-{\it{T}}$_c$ cuprates and many other layered compounds, the very weak interlayer hopping leads to quasi-two-dimensional electronic states in Sr$_2$TiO$_4$. Upon
electron doping (La substitution for Sr), the $d_{xy}$ states should first form a two-dimensional cylindrical Fermi surface. 
X-ray diffaction showed no impurity phases and rocking curves with a narrow full width at half maximum (FWHM) of 12.3 arcseconds and 28.6 arcseconds for the 5\% and 15\% La-doped Sr$_2$TiO$_4$ films, respectively. The resistivity of doped  films was found to be comparable with previous thin films grown by MBE~\cite{schlom1998searching}, but more than 4 orders of magnitude more conducting than previously reported bulk samples~\cite{Sugimoto1998315}. A negative Hall resistivity confirmed the presence of electron-like charge carriers. 

% ----------------Fermi surface map, two dimensional electronic structure, single dxy band --------------
ARPES measurements of La doped Sr$_{2}$TiO$_4$ reveal an electron-like band
near $E_F$ and a Fermi surface comprised of a single circular electron
pocket centered at $\Gamma$ [Fig.~\ref{fig:ARPES}(a)], in good agreement
with DFT calculations. The integrated density of states (DOS) up to higher binding
energies indicates a gap of  4 eV between bottom of the $d_{xy}$ band and the top of the O $2p$ valence band, consistent with earlier optical
measurements~\cite{Matsuno:2005hk}. Thus, the doped electrons indeed populate the $d_{xy}$ band instead of only forming deeply bound impurity states as has been observed in many doped, non-conducting oxides.  
It should be noted that our ARPES spectra show no clear sign of oxygen vacancies, whose signature is a non-dispersive in-gap state 1 eV below the bottom of the conduction band in SrTiO$_{3}$ \cite{Aiura200261,Meevasana:2011bh,MeevasanaNJP2010,SantanderSyro:2011hf}. Furthermore, we did not find any illumination dependence of the ARPES spectra, like that observed in SrTiO$_3$~\cite{SantanderSyro:2011hf, Meevasana:2011bh}.

%----------- EvsK and Franck-Condon Broadening---------------
A host of additional unusual spectral features are also observed in the ARPES spectra (Fig.~\ref{fig:analysis}), which
cannot be explained within a simple rigid band shift scenario. In addition to the 
low-lying electronic states crossing $E_F$, we also observe a broad spectral feature extending to higher binding energies. Second, the spectra exhibit a set of characteristic shoulders separated by 93 meV, which is also not present in the DFT calculations. Third, the measured carrier concentrations are substantially lower than the measured La doping levels. 

The Fermi wavevectors ($k_F$s) and the total Luttinger volume extracted from the Fermi surface maps [Fig.~\ref{fig:ARPES}(a)]  yield a doping level of 0.013 $\pm$ 0.004 electrons per Ti atom (corresponding to a sheet carrier density of 0.9 $\pm $ 0.3 $\times 10^{13}$ cm$^{-2}$ per TiO$_2$ layer) and 0.039 $\pm$ 0.005 electrons per Ti atom (2.6 $\pm 0.3 \times 10^{13}$ cm$^{-2}$ per TiO$_{2}$ layer) for the nominal 5 \% and 15 \% La-doped samples, respectively. The La content was confirmed by x-ray photoelectron spectroscopy to be $4.0\pm0.5\%$ and $14.4\pm0.5\%$, respectively [Fig.~\ref{fig:ARPES}(c), \ref{fig:ARPES}(d)]. The unexpectedly low carrier activation is reminiscent of the low free carrier activations measured at the LaAlO$_3$/SrTiO$_3$ interface~\cite{nakagawa2006some,Thiel:2006eo}. This could potentially be attributed to electrically inactive lanthanum
atoms associated with either dopant-induced disorder or vertically running SrO
faults, which have been observed by electron
microscopy~\cite{Lee:2013eg,tian2001transmission}. The degree of disorder can be
estimated from the lineshape of the momentum distribution curve (MDC) taken at
$E_F$ [Fig.~\ref{fig:analysis}(g) and \ref{fig:analysis}(h)]. Here, a Gaussian
lineshape is found to yield a good fit to the observed MDCs, resulting in
momentum width values of $\Delta k = 0.14~\mathrm\AA^{-1}$  and 0.18 $\mathrm
\AA^{-1}$ (FWHM), which is of the order of 1-2 unit cells. Understanding the nature of the disorder and how to increase the carrier mobility will be essential for future investigations into searching for possible superconductivity in doped or gated Sr$_{2}$TiO$_{4}$.

%%%%%%%%%%%%%%%%%%%%%%%%%%%%%%%%%%%FIG 2%%%%%%%%%%%%%%%%%%%%
\begin{figure}
\begin{center}
\includegraphics {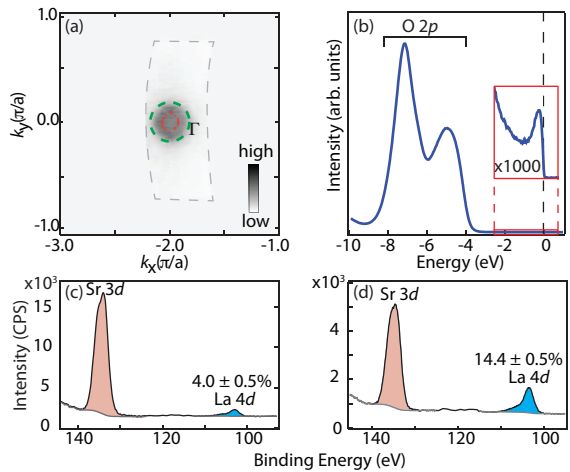}
\caption{(color online) (a) ARPES isoenergy intensity map with an integration window of $E_F$ +/- 10 meV of 5\% La-doped Sr$_2$TiO$_4$ shows a single circular pocket centered at the Brillouin zone center. The red marks indicate the extracted Fermi momenta via fits of momentum distribution curves (MDCs) and the green dashed line is the expected Fermi momenta. (b) Integrated EDC of the valence band indicates a gap of $\sim4$ eV. (c, d) XPS measurements reveal 4.0\% and 14.4\% lanthanum concentrations in nominal 5\% and 15\% doped films, respectively. 
% (d) MDC spectrum with fits to  a Lorentzian (green) and Gaussian (blue) spectral function. 
\label{fig:ARPES}}
\end{center}
\end{figure}

%%%%%%%%%%%%%%%%%%%%%%%%%%%%%%%%%%%FIG 2%%%%%%%%%%%%%%%%%%%%

Next, we focus on the 93 meV shoulders observed in our ARPES measurements, where peaks in the spectra show an unusual quantization of the electronic states  (Fig.~\ref{fig:analysis}). One possible scenario would be the formation of quantum well (QW) states as observed in Si-doped $\beta$-Ga$_2$O$_3$~\cite{iwaya2011atomically,Richard:2012fu}, where the impurity potential strongly confines the electron states in three-dimensions around the dopant La atom, resulting in the
formation of quantized levels. In this QW picture the energy
separation arising from the dopant ions should be inversely proportional to the impurity-impurity distance,
which should decrease with increasing lanthanum concentration. Our ARPES measurements
of a much more heavily doped (nominal 15\%) La-doped Sr$_2$TiO$_4$ film show a smaller mean free
path and larger $k_F$, but yield an essentially identical characteristic energy separation
($93\pm7$ meV) [Fig.~\ref{fig:analysis}(f)]. Furthermore, the states reported in Si-doped $\beta$-Ga$_2$O$_3$ are essentially dispersionless unlike the bands in Fig. 3(a) and 3(b). Therefore, the QW interpretation cannot fully account for the quantization of the quasi-localized
states observed in Sr$_2$TiO$_4$. In addition, we have considered the possibility of QW states due to the finite thickness of the thin films (Supplemental Information), but due to the very weak interlayer hopping along the $c$ axis for the $d_{xy}$ orbitals, no appreciable subband formation can be observed for the relevant $d_{xy}$ bands, only for the unoccupied $d_{xz}$ and $d_{yz}$ states.

Another possible scenario would be that of a bosonic mode strongly coupled to the electronic states resulting in satellite peaks separated by the
characteristic boson energy.  An infrared-active $E_u$ (LO) phonon mode
at 727 cm$^{-1}$ (90 meV) has been reported in
Sr$_2$TiO$_4$~\cite{PhysRevB.37.3381, Fennie2003}, involving the motion of oxygen anions creating an in-plane dipole which
can interact with charge carriers. We note that similar phonon replicas with comparable energies (100 meV) have also recently been reported in monolayer FeSe thin films grown on SrTiO$_{3}$ which have been argued to be relevant to the enhanced superconducting transition temperature \cite{lee2014}. This suggests that Sr$_{2}$TiO$_{4}$ may also be a potentially interesting substrate for monolayer FeSe, in addition to investigations of doped Sr$_{2}$TiO$_{4}$ itself. A pure el-ph coupling scenario alone, however,
cannot fully explain the broadened spectra
and the substantially reduced Luttinger volumes observed in our ARPES
measurements.

%%%%%%%%%%%%%%%%%%%%%%%%%%%%%%%%%%%FIG 3%%%%%%%%%%%%%%%%%%%%
\begin{figure}
\begin{center}
\includegraphics {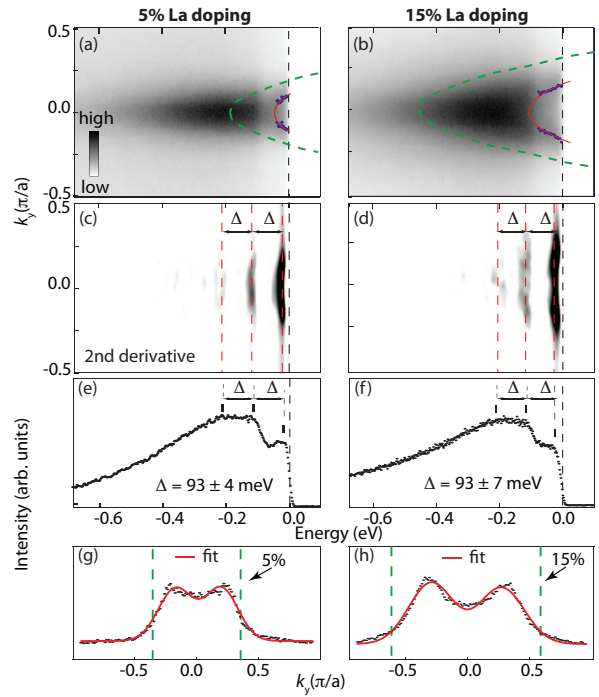}
\caption{(a, b) ARPES spectra taken along the $(0,0) - (0,\pi)$ direction of the 5\% and 15\% La-doped Sr$_2$TiO$_4$ films. The blue symbols mark the peak positions from Gaussian fits to the MDCs. The green and red curves are DFT calculated bands and parabolic fits to the peak positions, respectively. (c, d) The second derivative plots and (e, f) the EDCs show quantized states with a periodic separation of 93 meV for both films. The EDCs are integrated $\pm$0.01{\AA} around  Fermi momenta and the shifts from $E_F$ could be due to the experimental resolution.  (g, h) The Gaussian fits to the MDCs reveal FWHMs of 0.14 \AA$^{-1}$ and 0.18 \AA$^{-1}$ of the 5\% and 15\% doped samples, respectively. 
\label{fig:analysis}}
\end{center}
\end{figure}

%%%%%%%%%%%%%%%%%%%%%%%%%%%%%%%%%%%FIG 3%%%%%%%%%%%%%%%%%%%%

% -------------------Theory model calculations---------------------

In order to provide a theoretical understanding of the observed spectra, we consider a model with tight-binding
conduction electrons interacting with dispersionless phonon modes and disorder as described by the Hamiltonian~\cite{Mahan}

\begin{eqnarray}
H\!&=&\! -\sum_{i,j}t_{ij}\left(c^{\dagger}_{i}c_{j}+ {\rm h.c.}
 \right)-\sum_{i,j}g_{i,j}c_{i}^{\dagger}c_{i}(a_{j}+a_{j}^{\dagger})\nonumber\\
\!\!\!\!\!\! &&+\omega_0\sum_{i}a_{i}^{\dagger}a_{i} +\sum_{i}\xi_{i}c^{\dagger}_{i}c_{i},
\label{HH}
\end{eqnarray}

% {\color{blue}The electron-phonon (el-ph) polar coupling is described by a non-local coupling $g_{i,j}$ and the relevant phonon mode is the aforementioned LO infrared-active polar $E_u$ mode ~\cite{PhysRevB.37.3381, Fennie2003}.} 

%%%%%%%%%%%%%%%%%%%%%%%%%%%%%%%%%%%FIG 4%%%%%%%%%%%%%%%%%%%%
\begin{figure}
\begin{center}
\includegraphics {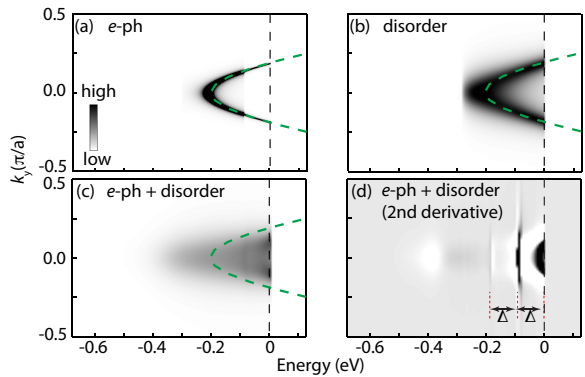}
\caption{Calculated ARPES spectra considering (a) el-ph coupling for $\lambda=0.15$ only ($\lambda$ is defined by equation (4) in Supplementary), (b) disorder only, and (c) both el-ph coupling and disorder, showing the important interplay of el-ph coupling and disorder.  (d) The second derivative of (c), demonstrating the quantization of quasi-localized electronic states. 
\label{fig:Theory}}
\end{center}
\end{figure}

%%%%%%%%%%%%%%%%%%%%%%%%%%%%%%%%%%%FIG 4%%%%%%%%%%%%%%%%%%%%

\noindent where $c_{i}^{\dagger}(c_{i})$ creates (destroys) an electron at site $i$, and
$a_{i}^{\dagger}(a_{i})$ creates (destroys) a phonon of frequency $\omega_0$ at
site $i$. The hopping matrix elements $t_{ij}$, fitted from DFT calculations,
connect neighboring sites up to fifth neighbors. In order to provide the simplest theoretical model to account for experimental evidences, the el-ph polar coupling is described by a non-local coupling $g_{i,j}$ and the relevant phonon mode is the
aforementioned LO infrared-active polar $E_u$ mode at a frequency $\omega_0$ of 
90 meV in Sr$_2$TiO$_4$~\cite{PhysRevB.37.3381, Fennie2003}.
The last term is an
Anderson term~\cite{Mahan}, where $\xi_i$ are energy levels distributed
according to a bimodal probability
$P(\xi)=(1-x)G_{\sigma}(\xi)+xG_{\sigma}(\xi-\xi_{b})$, where $G(\xi)$ is a Gaussian
function of variance $\sigma$, with the disorder level
$\xi_b$ extrapolated from supercell DFT calculations and $\sigma = 60$ meV. 
To investigate the role of disorder, we performed supercell DFT
calculations for five different configurations of lanthanum atoms [see Supplementary
Fig. 1(a)-1(f)], and weighting the different configurations, we find the growth of an impurity band localized
around -0.3 eV which merges into the bottom of the coherent Ti $d_{xy}$ band.

To determine the spectral function from the Hamiltonian in Eq.~(\ref{HH}), we
use a combined Coherent Potential Approximation (CPA)~\cite{Soven1967} and
phonon-phonon Non-Crossing Approximation (PPNCA)~\cite{Engelsberg1963,Cappelluti2003} to treat
both the local disorder and non-local el-ph coupling. The details of the formalism
are reported in the supplementary and in a theoretical paper~\cite{Theory}. As output
of our calculations, we obtain the self-energy $\sum(k, \omega)$, and thus
the spectral function $A(k,\omega) = (-1/\pi)\Im G(k, \omega)$. It is clear that considering only the el-ph
coupling [Fig.~\ref{fig:Theory}(a)] or disorder alone [Fig.~\ref{fig:Theory}(b)] cannot qualitatively describe the experimental ARPES
spectra. The former is missing the broad structure below
- 0.1 eV and the latter is missing the observed quantized phonon
signatures. In contrast, including both el-ph coupling and
disorder [Fig.~\ref{fig:Theory}(c)], the spectral function shows the quantized
levels (shown clearly in the second derivative in Fig.~\ref{fig:Theory}(d), as well as the broad and intense spectral weight at
higher binding energy, even when the value of the el-ph coupling strength is not
large (see Supplementary for additional details). 
Inclusion of the impurities' spatial correlations, neglected in this simple approach, 
could lead to a larger shift of the broad spectral weight to higher binding energy  \cite{DCA}. 
The apparent discrepancy between the nominal lanthanum concentration and the small Luttinger volume can be attributed to the presence of strong disorder in the system, consistent with earlier calculations which also considered the effect of strong disorder on the low energy electronic structure~\cite{haverkort2011electronic,berlijn2012transition,berlijn2012effective}. On the other hand, in our case, impurity induced disorder alone is not enough to explain such a large deviation of the nominal doping from the Luttinger volume
as is seen from the comparison of Fig. 4 b) and c). The reason for such
a dramatic effect of the e-ph interaction in conjunction with a strongly disordered energy
landscape originates from an increase in the binding energy of the states, which is caused by the e-ph interaction. By pushing the electronic states to higher energies and reducing the density of states near E$_F$, we find a strongly non-linear and non-perturbative decrease in the Luttinger volume and hence, a drop in the carrier density, particularly at the impurity sites~\cite{Theory} (see also Supplemental Information). 

%\textbf{Discussion}

In summary, our investigation of an epitaxially stabilized 2DEL in La-doped Sr$_{2}$TiO$_{4}$ reveals a single electronic band of $d_{xy}$ character crossing the Fermi level.
These electronic states, however,  are strongly modified due to the interplay
between dopant-induced disorder and el-ph coupling, resulting in
quantized semi-localized electronic states, as recently observed in monolayer FeSe on SrTiO$_{3}$~\cite{ge2014superconductivity}, and a striking reduction of the expected Luttinger volume. Understanding the nature of the disorder and the interplay with the electron-phonon coupling will be essential for increasing the carrier mobility and realizing superconductivity in doped Sr$_{2}$TiO$_{4}$. Since Sr$_2$TiO$_4$ is less prone to forming oxygen vacancies~\cite{Sugimoto1998315}, it should provide an intriguing new platform for controllable investigations of exotic superconductivity and electron-phonon interactions in oxide heterostructures and interfaces. We have directly examined, for the first time, the interplay between disorder and el-ph coupling in a complex oxide, which we expect to be important in understanding the phenomenology of correlated transition metal oxides, where electron-phonon coupling and disorder are known to play important roles in the underlying physics, such as the 2DEL at the LaAlO$_3$/SrTiO$_3$ interface~\cite{nakagawa2006some,KalabukhovPRB2007,herranz2007high,SiemonsPRL2007, willmott2007structural,reinle2012tunable,Golalikhani2013,warusawithana2013laalo3}, where the observed carrier density is much less than the expected 0.5 electrons per Ti. Our work suggests that the interplay between disorder and el-ph coupling is one natural mechanism that could explain the unusually low carrier densities at complex oxide interfaces. 
}
\\
\
\\

This work was primarily supported by the Air Force Office of Scientific Research ((rants No. FA9550-12-1-0335 and No. FA9550-11-1- 0033) and the National Science Foundation (DMR-0847385) and through the Materials Research Science and Engineering Centers program (No. DMR-1120296, the Cornell Center for Materials Research). This work was performed in part at the Cornell NanoScale Facility, a member of the National Nanotechnology Infrastructure Network, which is supported by the National Science Foundation (Grant ECCS-0335765). Y.N. acknowledges support from the State Key Program for Basic Research of China (Grant No. 2015CB654901). M.U. acknowledges support from the Japanese Society for the Promotion of Science. S. Ciuchi and D.D.S.  acknowledge the computational support from the CINECA supercomputing center under the grant "ConvR\_aq\_caspF" and "IscrC\_DISCOX". D. D. S. acknowledges support from the CARIPLO Foundation through the MAGISTER project Rif. 2013-0726. S. C. acknowledges support PRIN project Rif. 2012X3YFZ2\_006.

\bibliographystyle{apsrev}
%\bibliographystyle{naturemag}
%\textbf{References}
%\bibliography{sto214.bib} % Produces the bibliography via BibTeX.

%\newpage

\end{document}